\documentclass[aps,prl,reprint,amsmath,amssymb,superscriptaddress]{revtex4-2}
\usepackage[svgnames,psnames]{xcolor}
\usepackage{graphicx,hyperref}
\hypersetup{
colorlinks=true,
citecolor=NavyBlue,
linkcolor=DarkRed
}
\usepackage{bm,ulem}

\begin{document}

\title{Theory of Huge Thermoelectric Effect Based on Magnon Drag Mechanism: Application to Thin-Film Heusler Alloy }

\author{Hiroyasu \sc{Matsuura} }
\email{matsuura@hosi.phys.s.u-tokyo.ac.jp}
\affiliation{Department of Physics, University of Tokyo, 7-3-1 Hongo, Bunkyo, Tokyo 113-0033, Japan}
\author{Masao \sc{Ogata}}
\affiliation{Department of Physics, University of Tokyo, 7-3-1 Hongo, Bunkyo, Tokyo 113-0033, Japan}
\affiliation{Trans-scale Quantum Science Institute, University of Tokyo, Bunkyo-ku, Tokyo 113-0033, Japan} 
\author{Takao \sc{Mori}}
\affiliation{International Center for Materials, Nanoarchitectonics (WPI-MANA), National Institute for Materials Science (NIMS), Tsukuba, Japan}
\affiliation{Graduate School of Pure and Applied Sciences, University of Tsukuba, Tennodai 1-1-1, Tsukuba 305-8671, Japan}
\author{Ernst \sc{Bauer}}
\affiliation{Institute of Solid State Physics, Technische Universit\"{a}t Wien, Vienna, Austria}

\date{\today}

\begin{abstract}
To understand the unexpectedly high thermoelectric performance observed in the thin-film Heusler alloy Fe$_2$V$_{0.8}$W$_{0.2}$Al, we study the magnon drag effect, generated by the tungsten based impurity band, as a possible source of this enhancement, in analogy to the phonon drag observed in FeSb$_2$. 
Assuming that the thin-film Heusler alloy has a conduction band integrating with the impurity band, originated by the tungsten substitution, we derive the electrical conductivity $L_{11}$ based on the self-consistent t-matrix approximation and the thermoelectric conductivity $L_{12}$ due to magnon drag, based on the linear response theory, and estimate the temperature dependent electrical resistivity, Seebeck coefficient and power factor.
Finally, we compare the theoretical results with the experimental results of the thin-film Heusler alloy to show that the origin of the exceptional thermoelectric properties is likely to be due to the magnon drag related with the tungsten-based impurity band.    
\end{abstract}

\maketitle

\textit{Introduction.}---
Thermoelectric materials have attracted much attention because they can directly convert thermal energy to electric energy~\cite{Mori_Koumoto,Bell,Petsagkourakis}. 
Especially, the development of thermoelectric materials, utilizing magnetism, has been in the focus, and many materials with high thermoelectric performance have been found~\cite{Fahim,Tsujii,Acharya,Zheng,Vaney}. 
The efficiency of the thermoelectric conversion is expressed by the figure of merit, $ZT$, defined by $ZT \equiv S^2\sigma T/\kappa $ where  $S$, $\sigma$, $T$ and $\kappa$ are the Seebeck coefficient, electrical conductivity, temperature and thermal conductivity, respectively.
However, it is well known that $ZT$ is usually much lower than unity, because it is difficult to control these physical quantities independently.

Recently, it was found that a thin-film Heusler alloy, Fe$_2$V$_{0.8}$W$_{0.2}$Al, shows a huge $ZT$ ($ZT \sim$ 5 ) at $T\sim$ 350\,K, deriving from a huge power factor defined as $PF \equiv S^2\sigma$~\cite{Hinterleitner}.
The origin of these huge $ZT$ and $PF$ is expected to be related to the anomalous temperature dependence of the electrical resistivity and the Seebeck coefficient, because the electrical resistivity changes from a metallic behavior to a semiconducting behavior at $T \sim 350$\,K, and the Seebeck coefficient has a peak structure with a huge value ($S \sim -500 $\,$\mu$V/K) around this temperature. 

In a previous study, on the basis of the first principles calculation, the origin of this huge Seebeck coefficient was suggested to be a result of the large mobility due to many Weyl points and a large logarithmic energy derivative of the electronic density of states near the Fermi energy~\cite{Hinterleitner}.
On the other hand, it was also claimed~\cite{Alleno} that the crystal structure assumed in ref. \cite{Hinterleitner}  is different from the experimental one.
Then, it was reported that a new alloy model suggested in ref. \cite{Alleno} gives only rise to a Seebeck coefficient $S \sim 30$\,$\mu$V/K at $T \sim 400$\,K , which is much smaller than the experimental value. 
However, the actual alloy structures of Fe$_2$V$_{0.8}$W$_{0.2}$Al have not been fully explored both theoretically and experimentally. Furthermore, a contribution of magnetism related to the thin-film Heusler alloy~\cite{Hinterleitner} to the Seebeck coefficient has not yet been taken into account.
In addition to recent experimental reports, revealing an enhancement of the Seebeck coefficient of various systems through magnetic interactions~\cite{Fahim,Acharya,Vaney}, it has recently been experimentally demonstrated that spin fluctuation enhances the Seebeck coefficient of a doped itinerant ferromagnetic Fe$_2$VAl system~\cite{Tsujii}.

The temperature dependencies of the electrical resistivity and Seebeck coefficient observed in this thin-film Heusler alloy are very similar to those in FeSb$_2$:
FeSb$_2$ shows a huge Seebeck coefficient at low temperatures ($T \sim 10$\,K ), and at the same temperature, the electrical resistivity changes its temperature dependence to the semiconducting behavior as the temperature decreases~\cite{Bentien}. 
The origin of this huge Seebeck effect observed in FeSb$_2$ has been suggested being caused by a phonon drag, in which acoustic phonons couple with large effective mass electrons in an impurity band~\cite{Battiato,Takahashi,Matsuura}.    
From the analogy with FeSb$_2$, the origin of huge Seebeck effect observed in the thin-film Heusler alloy is supposed to be magnon drag related in the context of an impurity band and the conduction band with a large effective electron masses. 

The contribution of the magnon drag to the Seebeck effect has been studied experimentally~\cite{Blatt,Trego,Grannemann,Watzman} and theoretically~\cite{Bailyn,Sugihara,Miura,Imai,Yamaguchi} from the 1960s.
However, it appears that the magnon drag, related with an impurity band such as for the present alloy, is not sufficiently understood. 

In this letter, we study the magnon drag effect with an impurity state to clarify the origin of the huge Seebeck coeffcient and $PF$ observed in the thin-film Heusler alloy. 
Firstly, since the electronic state of the thin-film Heusler alloy has not been entirely understood yet, we assume an electronic state from the view point of a dimensional reduction. 
Extending the phonon drag theory studied in FeSb$_2$~\cite{Matsuura} to the thin-film Heusler alloy, we study the temperature dependence of the electrical resistivity, Seebeck coefficient, and PF related with such an impurity state.
We then compare the obtained theoretical results with experimental results to understand the origin of huge thermoelectric effect observed in the thin-film Heusler alloy.

\begin{figure}[h]
\begin{center}
\rotatebox{0}{\includegraphics[angle=0,width=1\linewidth]{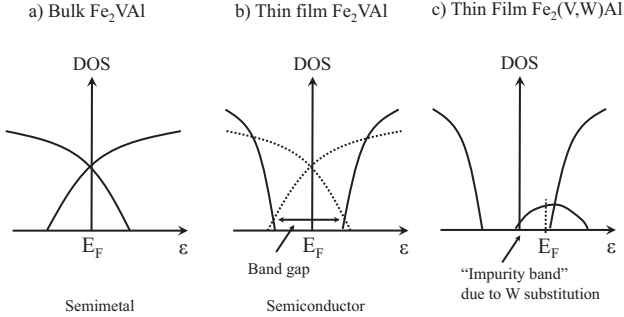}}
\caption{ Schematic picture on the electronic state of a) bulk Fe$_2$VAl, b) thin-film Fe$_2$VAl, and c) thin film Fe$_2$(V,W)Al.  }
\label{Fig1}
\end{center}
\end{figure}

\textit{Schematic picture of electronic states.}---
Firstly, we deduce the electronic state of the thin-film Heusler alloy based on the electronic state of bulk Fe$_2$VAl. 
Figure \ref{Fig1}(a) shows a schematic picture of the electronic state of the bulk Fe$_2$VAl near the Fermi level.
It is found~\cite{Hinterleitner,Singh,Weht} that this electronic state is a typical semimetallic state.   
In the thin-film, it is expected that the bandwidth decreases due to the dimensional reduction. 
Therefore, we suggest that a band gap appears by the lower dimension in the thin-film Fe$_2$VAl (Fig.\ref{Fig1}(b)).    
When vanadium (V) is replaced by tungsten (W) in this thin-film, it is natural to expect that impurity states appear near the bottom of the conduction band, because the energy level of 5d electrons in W is lower than the 3d energy level in V.
Figure \ref{Fig1}(c) shows a schematic picture of the electronic state of the thin-film Heusler alloy substituted by W.
In this letter, we study the electrical and thermal transports on the basis of this electronic state shown in Fig.\ref{Fig1}(c).

\textit{Model Hamiltonian and Formulation of Electric and Thermal transports.}---
To study the magnon drag based on the electronic state shown in Fig. 1(c), we use a following model Hamiltonian~\cite{Miura,Imai,Yamaguchi,Matsuura}.
\begin{eqnarray}
H = H_{\rm 0} + H_{\rm W} + H_{\rm mag} + H_{\rm e-mag},   \label{Hamil1}
\end{eqnarray}
where $H_{\rm 0}$, $H_{\rm W}$, $H_{\rm mag}$ and $H_{\rm e-mag}$ are Hamiltonians for a ferromagnetic conduction band, W sites, a ferromagnetic magnon, and an electron-magnon interaction, respectively.
These Hamiltonians are given as
$H_0 = \sum_{{\bf k}, \sigma} (\epsilon_{{\bf k}\sigma} -\mu) c_{{\bf k}\sigma}^{\dagger}c_{{\bf k}\sigma}$, 
$H_{W} =  V_0\sum_{\langle i \rangle } c_{i\sigma}^{\dagger}c_{i\sigma}$, 
$H_{mag} = \sum_{{\bf q}}\hbar \omega_{\bf q} b_{\bf q}^{\dagger}b_{\bf q}$, and
$H_{e-mag} = \frac{I}{\sqrt{V}}\sum_{{\bf k},{\bf q}}\bigr[ b_{\bf q}^{\dagger} c_{{\bf k}\uparrow}^{\dagger}c_{{\bf k} +{\bf q} \downarrow} + b_{\bf q} c_{{\bf k}+{\bf q}\downarrow}^{\dagger}c_{{\bf k}\uparrow}  \bigr]$,
where $c_{{\bf k}\sigma}$ or $c_{i\sigma}$ ($c_{{\bf k}\sigma}^{\dagger}$ or $c_{i\sigma}^{\dagger}$) is an annihilation (creation) operator of an electron with the wave number ${\bf k}$ on the $i$-th site and spin $\sigma =\uparrow\downarrow$; $b_{\bf q}$ ($b_{\bf q}^{\dagger}$) is an annihilation (creation) operator of a magnon with wave vector ${\bf q}$.
$\epsilon_{{\bf k}\sigma}$ is the energy dispersion in the ferromagnetic state, $\mu$ is a chemical potential,  $V_{0}$ is the strength of a random impurity potential, $\langle i \rangle$ is the position of impurities, and $\hbar \omega_{\bf q}$ is the energy dispersion of ferromagnetic magnons given by $\hbar\omega_{\bf q} = Dq^2$, where $D$ is the spin wave stiffness constant.
Finally, $I=J\sqrt{V}$ is the strength of the electron-magnon interaction, where $J$ and $V$ are the coupling constant between electron and magnon, and the volume of unit cell, respectively.
In this letter, we use the following simple energy dispersion: $\epsilon_{{\bf k}\uparrow}  = \frac{\hbar^2 k^2}{2m^*} - \Delta, \label{epsilon_k1}$ and $\epsilon_{{\bf k}\downarrow}  = \frac{\hbar^2 k^2}{2m^*}$,
where $m^*$ is the effective mass of conduction electrons, $\Delta$ is the energy difference between up spin and down spin electrons to express the ferromagnetic state, which corresponds to d orbitals of Iron (Fe) in FeV$_{0.8}$W$_{0.2}$Al~\cite{Hinterleitner}.
We assume that $\Delta$ is independent of temperature for simplicity.
Because the Fermi energy is located near the bottom of the conduction band or in the impurity band as shown in Fig. 1(c), the valence band is neglected although it will contribute at high temperatures.

To treat the random potential of the W site, we use a self-consistent $t$-matrix approximation~\cite{Saitoh,Ogata1,Yamamoto1,Matsuura,Matsubara}.
As discussed in Ref~\cite{Matsuura}, we define the retarded Green's function of electron with spin $\sigma$ as
\begin{eqnarray}
G^{R}_{\sigma}(k,\epsilon) = \frac{1}{\epsilon -\epsilon_{{\bf k}\sigma} -\Sigma^{R}_{\sigma}(\epsilon)  },\label{GF1}
\end{eqnarray}
where by the self-consistent t-matrix approximation, a retarded self-energy, $\Sigma^{R}_{\sigma}(\epsilon) $, is given as  
$\Sigma^{R}_{\sigma}(\epsilon) = \frac{n_i V_0}{1- \frac{V_0}{V}\sum_{\bf k}^\prime G^{R}_{\sigma}(k^\prime,\epsilon)  }.$  \label{sigma_self}  
Here, $n_i$ is the concentration of W sites.
The density of state (DOS) is obtained by 
$D_{\sigma}(\epsilon) = D_0 {\rm Im}[y_\sigma], $ 
where 
$D_0 = \frac{\sqrt{(m^*)^3 \epsilon_B}}{\sqrt{2}\pi^2 \hbar^3}, $
and $y_{\sigma}$ is determined by solving the cubic equation:
$y_{\sigma}^3 -2y_{\sigma} + \bigr( 1 + \frac{\epsilon + \Delta\delta_{\sigma,\uparrow}}{\epsilon_B} \bigr)y_{\sigma} - \nu = 0$ \cite{Matsuura}.
Here, $\nu \equiv 2\pi n_i \hbar^3/\sqrt{2(m^*)^3\epsilon_B^3}$. 
We assumed that $\epsilon_B$ ($\epsilon_B +\Delta$) is the binding energy of a single W impurity for down spin (up spin) as a first step.
It should be noted that the first principles calculation shows no spin splitting in 5d orbitals of W \cite{Hinterleitner}.

The Fermi energy ($E_F$) and the temperature dependence of the chemical potential are determined self-consistently by
$\sum_{\sigma}\int_{-\infty}^{\infty} f(\epsilon) D_{\sigma}(\epsilon) d\epsilon = \sum_{\sigma}\int_{-\infty}^{E_F}D_{\sigma}(\epsilon) d\epsilon = n_i, $
where $f(\epsilon)$ is the Fermi distribution function defined by $f(\epsilon) = 1/(e^{\beta(\epsilon-\mu)} + 1)$.


The electrical current (${\bf J}_{e}$) and the heat current due to electrons (${\bf J}_{Q}^{\it ele}$), and the heat current due to ferromagnetic magnons (${\bf J}_{Q}^{\it mag}$) are defined as
${\bf J}_{e} = e\sum_{{\bf k}\sigma}v_{{\bf k},\sigma} c_{{\bf k},\sigma}^{\dagger} c_{{\bf k},\sigma}$, $
{\bf J}_{Q}^{\it ele} = \sum_{{\bf k}\sigma}(\epsilon_{{\bf k}\sigma}  - \mu) v_{{\bf k},\sigma} c_{{\bf k},\sigma}^{\dagger} c_{{\bf k},\sigma}$, and ${\bf J}_{Q}^{\it mag} = \sum_{\bf q} \hbar \omega_{\bf q} \frac{\partial \omega_{\bf q}}{\partial q_x} b_{\bf q}^{\dagger}b_{\bf q}$, 
where $v_{{\bf k},\sigma} = \frac{1}{\hbar}\frac{\partial \epsilon_{{\bf k},\sigma}}{\partial k_x}$ and $e$ is the electron charge ($e <0$).

Under an electric field ${\bf E}$  and temperature gradient $\nabla T$,  the electrical current density ${\bf j}$ is described in the linear response theory as
${\bf j} = L_{11} {\bf E} + L_{12}\bigr( - \frac{{\bf \nabla} T}{T} \bigr)$, 
where $L_{11}$ and $L_{12}$ are electrical conductivity and thermoelectric conductivity, respectively~\cite{Behnia}.
These coefficients are calculated from the correlation function between the electrical currents, and that between the electrical and heat currents derived by Kubo and Luttinger~\cite{Kubo,Luttinger,Ogata2}:
\begin{eqnarray}
L_{ij} = \lim_{\omega \rightarrow 0} \frac{\Phi_{ij}(\omega + i\delta) - \Phi_{ij}(0)}{i\omega + i\delta},  \label{kL}
\end{eqnarray}
where $\omega$ is a frequency of the external field. 
In the present case, $L_{12}$ contains two components due to $J_{Q}^{\it ele}$ and $J_{Q}^{\it mag}$, which we refer to $L_{12}^{\it ele}$ and $L_{12}^{\it drag}$, respectively.

The transport coefficient $L_{11}$ due to the electrical currents and $L_{12}^{\it ele}$ owing to the electrical current and the heat current due to electrons are~\cite{Ogata2} 
\begin{eqnarray}
L_{11} &=&  \int d\epsilon \bigr(-\frac{\partial f(\epsilon)}{\partial \epsilon} \bigr) \sigma(\epsilon), \\
L_{12}^{\it ele} &=&  \frac{1}{e}\int d\epsilon \bigr(-\frac{\partial f(\epsilon)}{\partial \epsilon} \bigr) (\epsilon- \mu)\sigma(\epsilon), 
\end{eqnarray}
where $\sigma(\epsilon)$ is the function of electrical conductivity, depending on $\epsilon$.
The relaxation time of electrons is included in $\sigma(\epsilon)$. 
When we use a Green's function, which is obtained in Eq.  (\ref{GF1}), $\sigma(\epsilon)$ is given by
\begin{eqnarray}
\sigma(\epsilon) = \sum_{\sigma} \frac{e^2\sqrt{m^*} }{12 \pi^2 \hbar^2 }\frac{(\sqrt{x_{\sigma}^2 + \Gamma_{\sigma}(\epsilon)^2 } + x_{\sigma})^{\frac{3}{2}}}{\Gamma_{\sigma}(\epsilon)},  \label{cond}
\end{eqnarray}
where $x_{\sigma}= \epsilon + \Delta \delta_{\sigma,\uparrow}  -{\rm Re}\Sigma_{\sigma}^{\rm R}(\epsilon)$ and $\Gamma_{\sigma}(\epsilon) = -{\rm Im}\Sigma_{\sigma}^{\rm R}(\epsilon)$, respectively.
It has to be noted that we consider only the effect of the random potential, given by the self-consistent t-matrix approximation, and neglect the effect of relaxation due to the electron-magnon interaction in the calculation of the electrical conductivity $L_{11}$.

Next, we study the correlation function between the electrical current and the heat current of magnons defined as $\Phi_{12}(\tau) = \frac{1}{V}\langle T_\tau [ {\bf j}_{e}(\tau) {\bf j}_{Q}^{\it mag}(0) ] \rangle$, 
where $\tau$ is an imaginary time and $T_\tau$ denotes the imaginary time ordering operator \cite{Imai,Yamaguchi}. 
By the second order perturbation on the exchange interaction based on the Green's function of electrons, Eq. (\ref{GF1}), the correlation function due to the magnon drag is obtained as  
\begin{eqnarray}
&&\Phi_{12}^{\rm drag}(\omega) = i\omega \frac{I^2e(m^*)^2 }{48\pi^3 \hbar^5 \Gamma_{\rm mag}(T)} 
\int _{0}^{\epsilon^{\rm cut}_q} d\epsilon_{q} \frac{\beta \epsilon_{q} e^{\beta \epsilon_q}}{(e^{\beta \epsilon_{q}} -1)^2}  \nonumber \\
&& \times \int dx f(x) \biggr[ \frac{L_1^{+}}{\Gamma_{\downarrow}(x + \epsilon_{q})} - \frac{L_2^{+}}{\Gamma_{\downarrow}(x)} - \frac{L_1^{-}}{\Gamma_{\uparrow}(x)} + \frac{L_2^{-}}{\Gamma_{\uparrow}(x - \epsilon_{q})}\biggr],  \nonumber \\
&&
\label{magnon_drag}
\end{eqnarray}
where $\Gamma_{\rm mag}(T)$ and $\epsilon^{\rm cut}_q$ are the temperature dependent magnon relaxation rate, and an energy cutoff of magnons, respectively; $L_{1}^{\pm} = \epsilon_{q}  \pm \alpha\epsilon_{q} - \Delta - {\rm Re}\Sigma_{\downarrow}^{({\rm R})}(x + \epsilon_{q}) +  {\rm Re}\Sigma_{\uparrow}^{({\rm R})}(x)$ and $L_{2}^{\pm} = \epsilon_{q} \pm \alpha\epsilon_{q} - \Delta + {\rm Re}\Sigma_{\uparrow}^{({\rm R})}(x -\epsilon_{q}) -  {\rm Re}\Sigma_{\downarrow}^{({\rm R})}(x)$
where $\alpha =  \frac{\hbar^2 }{2m^* D} $.
In the supplemental material, we show the derivation of Eq. (\ref{magnon_drag}) in detail. 
Using Eq. (\ref{kL}), the thermoelectric conductivity due to the magnon drag, $L_{12}^{\rm drag}$, is obtained.
The vetex corrections, which are neglected in this letter for simplicity, have been discussed in ref. \cite{Yamaguchi}. 
 
\begin{figure}[h]
\begin{center}
\rotatebox{0}{\includegraphics[angle=0,width=1\linewidth]{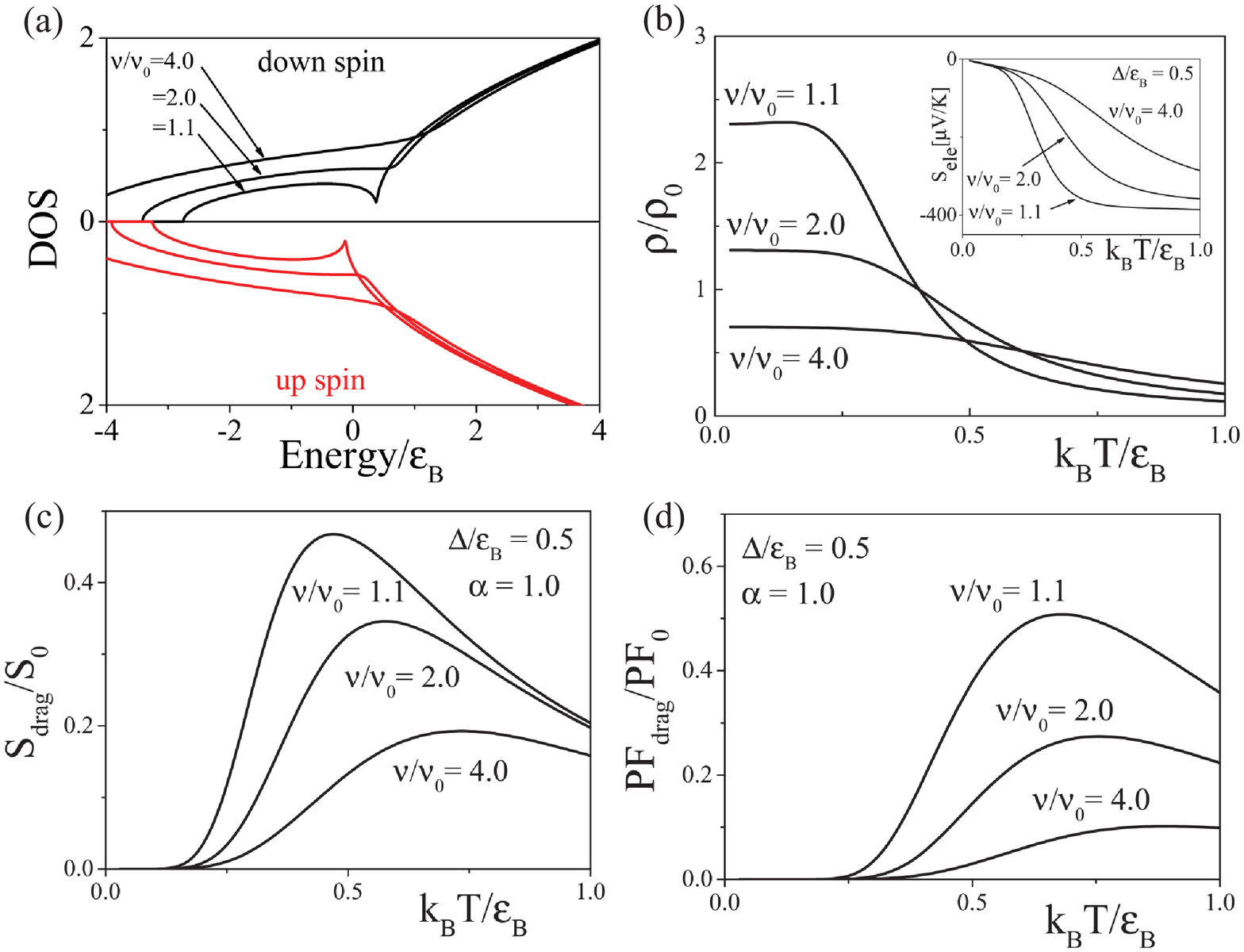}}
\caption{(a) Densitiy of states (DOS), (b) electrical resistivity, (c) Seebeck coefficient due to the magnon drag ($S_{\rm drag}$) and (d) the power factor due to the magnon drag (PF$_{\rm drag}$) for $\tilde{\nu} \equiv \nu/\nu_0 =1.1$, $2$ and $4$ and $\Delta/\epsilon_B = 0.5$. Inset of (b) Seebeck coefficient due to the heat current of electron ($S_{\rm ele}$). }
\label{Fig2s}
\end{center}
\end{figure}

\textit{Numerical Results.}---
Figure \ref{Fig2s}(a) shows the density of states for $\tilde{\nu} \equiv \nu/\nu_0 =1.1$, $2$ and $4$.
We set $\Delta/\epsilon_B = 0.5$.     
For $\tilde{\nu} =2$ and $4$, the impurity band hybridizes the conduction band naturally, while for $\tilde{\nu}=1.1$, the impurity band only slightly touches with the conduction band.      
The Fermi energy is located in $E_F/\epsilon_B \simeq -1.20$ for $\tilde{\nu} = 1.1$ , $-1.16$ for $\tilde{\nu} = 2$, and $-1.07$ for $\tilde{\nu} = 4.0$.
It should be noted that the chemical potential does not show a drastic temperature dependence.

Figure \ref{Fig2s}(b) shows the temperature dependent electrical resistivity ($\rho =1/L_{11}$) for $\tilde{\nu} =1.1$, $2$ and $4$. Here, $\rho_0 = 12 \pi^2 \hbar^2/\gamma e^2\sqrt{m^* \epsilon_B}  $.
It has to be noted that we introduce the dimensionless phenomenological parameter $\gamma$ to consider additional contributions of the valleys and other unspecified processes to the electrical conductivity.
As shown in Fig. \ref{Fig2s}(b), the resistivity increases gradually, as the temperature decreases from high temperatures, while around $k_BT/\epsilon_B \simeq 0.5 $, the resistivity drastically increases; the resistivity becomes constant at low temperatures.
This behavior is a result of the impurity band.  
We also conclude that the constant resistivity value at low temperatures depends on the impurity concentration.

Next, let us discuss the Seebeck coefficient due to the heat current of electrons, i.e. $S_{\rm ele} = L_{12}^{\rm ele}/TL_{11}$ and the Seebeck coefficient due to the magnon drag, i.e. $S_{\rm drag} = L_{12}^{\rm drag}/TL_{11}$. 
The inset of Figure \ref{Fig2s}(b) shows the temperature dependent Seebeck coefficients $S_{\rm ele}$ for $\tilde{\nu}=1.1$, $2$ and $4$.
As the impurity concentration decreases, the Seebeck coefficient increases, while the Seebeck coefficient does not show a peak structure. 
Figure \ref{Fig2s}(c) shows the temperature dependent term $S_{\rm drag}$, for $\tilde{\nu}=1.1$, $2$ and $4$ and $\alpha = 1.0$.
We assume a temperature dependent magnon relaxation rate, $\Gamma_{\rm mag} = \bigr(\hbar/ 2\tau_0 \bigr) T$, where $\tau_0$ is a constant.  
The factor $S_{0}$ is defined by $S_{0} = I^2(m^*)^{3/2}k_B^2 \tau_0/2e\pi\hbar^4 \sqrt{\epsilon_B}$. 
Note that the Seebeck coefficient does not depend on $\gamma$.
As shown in Fig.\ref{Fig2s}(c), the Seebeck coefficient increases as the impurity concentration decreases.
We also find that a peak structure of the temperature dependent Seebeck coefficient appears around $k_BT/\epsilon_B \simeq 0.5$  for $\nu/\nu_0 = 1.1$, $0.6$ for $\nu/\nu_0 = 2$ and $0.7$ for $\nu/\nu_0 = 4$. 
Figs. \ref{Fig2s}(d) show the temperature dependent power factor due to the magnon drag, $PF_{\rm drag}$, where we define $PF_{0} = S_{0}^2/\rho_0$.
The $PF_{\rm drag}$ traces closely the temperature dependent Seebeck coefficient as shown in Figs. \ref{Fig2s}(c) regarding several impurity concentrations, while we find that the peak temperature of $PF_{\rm drag}$ is higher than that of Seebeck coefficient, because of the distinct decrease of the electrical resistivity. 
It should be noted that the temperature dependences of $S_{\rm drag}$ and $PF_{\rm drag}$ are insensitive to $\alpha$, while these values strongly depend on $\alpha$ (See the supplemental material).

\textit{Discussion: Comparison with experiments.}---
Here, we compare the obtained theoretical results with the experimental results of the thin-film Heusler alloy.
Since there are no experimental data on theoretical parameters, we have chosen a set of the reasonable values: $\epsilon_B/k_B=300$\,K,  $m^{*}/m_0 = 10$, $J/k_B = 1000$\,K, $V =10^{-27}$\,$m^3$, and $\gamma = 10$.
It should be noted that the large effective mass is due to the large density of states of the conduction band as  shown from the first principles DFT calculations~\cite{Hinterleitner}; then, the impurity concentration ($n_i$) is of the order of $10 ^{27}$\,$m^{-3}$ for $\tilde{\nu}=1 \sim 4$, which is consistent with the concentration of W in Fe$_2$VAl.
Here, we set the life time of magnon as $\tau = \tau_0/T \sim 10^{-14}$\,s at $T=300$\,K.
This value is reasonable for a ferromagnetic metal~\cite{Zhang}.

\begin{figure}[h]
\begin{center}
\rotatebox{0}{\includegraphics[angle=0,width=1\linewidth]{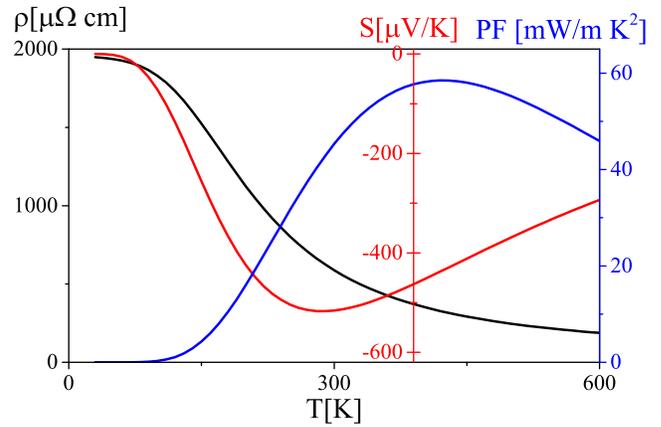}}
\caption{Temperature dependences of electrical resistivity, Seebeck coefficient due to the magnon drag, and power factor (PF) in the realistic parameters. }
\label{Fig3}
\end{center}
\end{figure}

Using these parameters, $\nu/\nu_0 = 4$ and $\alpha = 1.0$, the temperature dependent electrical resistivity and Seebeck coefficient due to the magnon drag,  as well as the power factor are displayed in Fig. \ref{Fig3}.
We find that the electrical resistivity attains $\rho \simeq$ 1000\,$\mu\Omega$cm at $T \sim 300$\,K; we also find that the Seebeck coefficient due to the magnon drag exhibits a peak structure, with $S_{\rm max} \sim -500$\,$\mu$V/K at $T \sim$ 300\,K.
The power factor reaches $PF \sim$ 60\,mW/m K$^2$ around $T \sim 400$\,K.
Since these theoretical results are similar to the experimental results, we presume that the origin of the huge Seebeck coefficient and the large $PF$ observed experimentally for the thin-film Heusler alloy is likely due to a magnon drag, related to the tungsten-based impurity band.

Finally, we comment on the life time of magnons.
In this letter, we used a simple temperature dependent life time of magnons.
However, the life time is expected to be very complicated in a real material, because it is derived from many kinds of scattering mechanisms such as impurity scattering, magnon-electron, magnon-magnon, magnon-phonon interactions related with (without) the Umklapp process, and so on.
The understanding of these microscopic mechanisms for the life time of magnon is a future problem.

\textit{Conclusion.}---
We studied the origin of the large Seebeck coefficient and unprecedented large PF observed in the thin-film Heusler alloy FeV$_{0.8}$W$_{0.2}$Al on the basis of the linear response theory.
Assuming that this thin-film alloy has a conduction band integrating with the impurity band originated from the W substitution, and by extending the microscopic phonon drag theory observed in FeSb$_2$, we derived $L_{11}$ based on the self-consistent t-matrix approximation and $L_{12}$ due to the magnon drag.
As a result, we found that the theoretical results of the Seebeck coefficient and PF are in agreement with the experimental ones.
Therefore, we concluded that the origin of these striking thermoelectric properties is likely due to the magnon drag related with the W-based impurity band.

\textit{Acknowledgments.}---
This work is supported by Grants-in-Aid for Scientific Research from the Japan Society for the Promotion of Science (No.\ JP18H01162, No.\ JP18K03482, and No.\ JP20K03802), and JST-Mirai Program Grant (No.\ JPMJMI19A1).


\begin{thebibliography}{9}                                                       
\bibitem{Bell}
L.\ E.\ Bell,  Science, {\bf 321}, 1457 (2008).
\bibitem{Mori_Koumoto}
K.\ Koumoto and T.\ Mori, {\it Thermoelectric Nanomaterials, Springer Series in Materials Science}, {\bf 182}, (2013).
\bibitem{Petsagkourakis}
I.\ Petsagkourakis et al., Sci. Tech. Adv. Mater., {\bf 19}, 836-862 (2018).
\bibitem{Fahim}
A.\ Fahim, N.\ Tsujii, and T.\ Mori: J.\ Mater.\ Chem.\ A {\bf 5}, 7545 (2017).
\bibitem{Tsujii}
N.\ Tsujii, A.\ Nishide, J.\ Hayakawa, and T.\ Mori, Science Advances, {\bf 5}, eaat5935 (2019).
\bibitem{Acharya}
S.\ Acharya, S.\ Anwar, T.\ Mori and A.\ Soni, J.\ Mater.\ Chem.\ C, {\bf 6}, 6489 (2018). 
\bibitem{Zheng}
Y.\ Zheng et al., Sci.\ Adv.\ {\bf 5}, eaat9461 (2019).
\bibitem{Vaney}
J.\ B.\ Vaney, S.\ A.\ Yamini, H.\ Takaki, K.\ Kobayashi, N.\ Kobayashi, and T.\ Mori, Mater.\ Today Phys., {\bf 9}, 100090 (2019).
\bibitem{Hinterleitner}
B.\ Hinterleitner, et al., Nature, {\bf 576}, 85 (2019).
\bibitem{Alleno}
E.\ Alleno, et al., Phys. Chem. Chem. Phys. , {\bf 22}, 22549 (2020).
\bibitem{Bentien}
A.\ Bentien, et al., Europhys. Lett. {\bf 80}, 17008 (2007). 
\bibitem{Battiato} 
M.\ Battiato, J.\ M.\ Tomczak, Z.\ Zhong, and K.\ Held, Phys.\ Rev.\ Lett. , {\bf 114}, 236603 (2015).
\bibitem{Takahashi}
H.\ Takahashi, et al., Nat. Commun. {\bf 7}, 12732 (2016). 
\bibitem{Matsuura}
H.\ Matsuura, et al., J. Phys. Soc. Jpn.  {\bf 88}, 074601 (2019).
\bibitem{Blatt}
J.\ Blatt, et al., Phys.\ Rev.\ Lett.\ {\bf 18}, 395 (1967).
\bibitem{Trego}
A.\ L.\ Trego and A.\ R.\ Mackintosh, Phys.\ Rev.\ {\bf 166}, 495 (1968).
\bibitem{Grannemann}
G.\ N.\ Grannemann and L.\ Berger., Phys.\ Rev.\ B {\bf 13}, 2072 (1976).
\bibitem{Watzman}
S.\ Warzman, et al., Phys. Rev. B {\bf 94}, 144407 (2016). 
\bibitem{Bailyn}
M.\ Bailyn, Phys.\ Rev.\ {\bf 126}, 2040 (1962).  
\bibitem{Sugihara}
K.\ Sugihara, J.\ Phys.\ Chem.\ Solids {\bf 33}, 1365 (1972).
\bibitem{Miura}
D.\ Miura and A.\ Sakuma, J. Phys. Soc. Jpn. {\bf 81} 113602 (2012).
\bibitem{Imai}
Y.\ Imai and H.\ Kohno, J. Phys. Soc. Jpn. {\bf 87} 073709 (2018).
\bibitem{Yamaguchi}
T.\ Yamaguchi, H.\ Kohno, and R.\ Duine, Phys.\ Rev.\ B {\bf 99} 094425 (2019).
\bibitem{Singh}
D.\ J.\ Singh, and I.\ I.\ Mazin, Phys.\ Rev.\ B {\bf 57} 14352 (1998).
\bibitem{Weht}
R.\ Weht, and W.\ E.\ Pickett, Phys.\ Rev.\ B {\bf 58} 6855 (1998).   
\bibitem{Saitoh} 
M.\ Saitoh, H.\ Fukuyama, Y.\ Uemura, and H.\ Shiba, J.\ Phys.\ Soc.\ Jpn. {\bf 27}, 26 (1969).
\bibitem{Yamamoto1} 
T.\ Yamamoto and H.\ Fukuyama, J.\ Phys.\ Soc.\ Jpn. {\bf 87}, 024707 (2018).
\bibitem{Ogata1} 
M.\ Ogata and H.\ Fukuyama, J.\ Phys.\ Soc.\ Jpn. {\bf 86}, 094703 (2017).
\bibitem{Matsubara} 
M.\ Matsubara, K.\ Sasaoka, T.\ Yamamoto and H.\ Fukuyama, J.\ Phys.\ Soc.\ Jpn. {\bf 90}, 044702 (2021).
\bibitem{Behnia}
Kamuran Behnia, {\it Fundamentals of Thermoelectricity}, (Oxford University press, Oxford, 2015).
\bibitem{Kubo}
 R. Kubo, J.\ Phys.\ Soc.\ Jpn. {\bf 12}, 570 (1957).
\bibitem{Luttinger} 
J.\ M.\ Luttinger, Phys.\ Rev.\ {\bf 135}, A1505 (1964).
\bibitem{Ogata2} 
M.\ Ogata and H.\ Fukuyama, J.\ Phys.\ Soc.\ Jpn. {\bf 88}, 074703 (2019).
\bibitem{Zhang} 
Y. Zhang et al. Phys.\ Rev.\ Lett.\ {\bf 109}, 087203 (2012). 
 
\end{thebibliography}
\end{document}


\widetext
\clearpage
\begin{center}
  \textbf{\large
Supplemental Materials
  }
\end{center}
\setcounter{section}{0}
\setcounter{equation}{0}
\setcounter{figure}{0}
\setcounter{table}{0}
\setcounter{page}{1}
\makeatletter
\renewcommand{\thesection}{S-\Roman{section}}
\renewcommand{\theequation}{S\arabic{equation}}
\renewcommand{\thefigure}{S\arabic{figure}}
\renewcommand{\thetable}{S\arabic{table}}
\renewcommand{\bibnumfmt}[1]{[S#1]}
\renewcommand{\citenumfont}[1]{S#1}
\section{I: Derivation of Eq. (7)}
In this supplemental material, we derive eq.\ (7) on the basis of the linear response theory. 
By the second order perturbation of $H_{\rm e-mag}$, the lowest order correlation function is given as
\begin{eqnarray}
\Phi_{\rm 12}({\bf q}=0, \tau,0)  = &&\frac{1}{2V}\int_{0}^{\beta} d\tau_1 \int_{0}^{\beta} d\tau_2 \langle T_{\tau} \bigr[ H_{\rm e-mag}(\tau_1)H_{\rm e-mag}(\tau_2)j_{e}(\tau)j_Q^{mag}(0) \bigr] \rangle.
\end{eqnarray}
By Wick's theorem, $\langle T_{\tau} \bigr[ \cdots \bigr] \rangle$ is expressed as
\begin{eqnarray}
&&\langle T_{\tau} \bigr[ H_{\rm e-mag}(\tau_1)H_{\rm e-mag}(\tau_2)j_{e}(\tau)j_Q^{mag}(0) \bigr] \rangle = 
\frac{I^2}{V}\sum_{{\bf k}, {\bf q}} \frac{e\hbar k_x}{m}\frac{2D^2q^2q_x}{\hbar} \nonumber \\
&&\times \bigr[ G_{\uparrow}({\bf k}, \tau -\tau_1)G_{\downarrow}({\bf k}+{\bf q}, \tau_1 -\tau_2) G_{\uparrow}({\bf k}, \tau_2 -\tau)D({\bf q},-\tau_1)D({\bf q},\tau_2) \nonumber \\
&& + G_{\downarrow}({\bf k}, \tau_1 - \tau )G_{\uparrow}({\bf k}, \tau -\tau_2) G_{\uparrow}({\bf k}-{\bf q}, \tau_2 -\tau_1)D({\bf q},-\tau_1)D({\bf q},\tau_2) \nonumber \\
&&G_{\uparrow}({\bf k}-{\bf q}, \tau_1 -\tau_2)G_{\downarrow}({\bf k}, \tau_2 -\tau) G_{\downarrow}({\bf k}, \tau -\tau_1)D({\bf q},\tau_1)D({\bf q},-\tau_2)  \nonumber \\
&& + G_{\downarrow}({\bf k}+{\bf q}, \tau_2 - \tau_1 )G_{\uparrow}({\bf k}, \tau_1 -\tau) G_{\uparrow}({\bf k}, \tau -\tau_2)D({\bf q},\tau_1)D({\bf q},-\tau_2) \bigr],
\end{eqnarray}
where the Green's functions for electron and magnon are defined as
\begin{eqnarray}
G_{\sigma}({\bf k}, \tau) &=& -\langle T_{\tau} [c_{k\sigma}(\tau)c_{k\sigma}^{\dagger}(0)] \rangle, \\ 
D ({\bf q}, \tau) &=& -\langle T_{\tau} [b_{q}(\tau)b_{q}^{\dagger}(0)] \rangle.
\end{eqnarray}

By the Fourier transformations of 
\begin{eqnarray}
\Phi_{\rm 12}(i\omega_\lambda) &=& \int d\tau e^{i\omega_\lambda \tau}\Phi_{12}(\tau), \\
G_{\sigma}({\bf k}, \tau) &=& \frac{1}{\beta} \sum_{n} e^{-i\epsilon_{n} \tau} G_{\sigma}({\bf k}, \tau), \\ 
D({\bf q}, \tau) &=& \frac{1}{\beta} \sum_{m} e^{-i\omega_{m} \tau} D({\bf q}, \tau),  
\end{eqnarray}
where $\omega_\lambda$, $\epsilon_n$ and $\omega_m$ are the Matsubara frequencies with $\omega_\lambda = 2\lambda\pi k_{\rm B}T$,  $\epsilon_n = (2n+1)\pi k_{\rm B}T$, and $\omega_m = 2m\pi k_{\rm B}T$ with $\lambda$, $n$ and $m$ being the integers,
the correlation function is given as
\begin{eqnarray}
\Phi_{\rm 12}(i\omega_\lambda) &=& \frac{I^2}{2V^2}\sum_{{\bf k}, {\bf q}} \frac{e\hbar k_x}{m}\frac{2D^2q^2q_x}{\hbar} \frac{2}{\beta^2} \bigr[  G_{\uparrow}({\bf k}, i\epsilon_n)G_{\downarrow}({\bf k}+{\bf q},i\epsilon_n + i\omega_{m} + i\omega_{\lambda} ) G_{\uparrow}({\bf k}, i\epsilon_{n} + i\omega_{\lambda}) \nonumber \\
&& + G_{\downarrow}({\bf k}, i\epsilon_n)G_{\uparrow}({\bf k}-{\bf q},i\epsilon_n - i\omega_{m} ) G_{\downarrow}({\bf k}, i\epsilon_{n} + i\omega_{\lambda})\bigr] D({\bf q},i\omega_{m})D({\bf q},i\omega_{m} + i\omega_{\lambda}). \nonumber \\
&& \label{phi12_A1}
\end{eqnarray}
Here, the thermal Green's functions for electron and magnon are defined as
\begin{eqnarray}
G_{\sigma}({\bf k}, i\epsilon_n) &=& \frac{1}{i\epsilon_{n} - (\epsilon_{{\bf k},\sigma} -\mu) - {\rm sgn}(\epsilon_n)\Sigma(\epsilon_n)}, \\
D({\bf q},i\omega_{m}) &=& \frac{1}{i\omega_{m} - \hbar \omega_{\bf q} -{\rm sgn}(\omega_m)\Gamma_{\rm mag}(T) }.
\end{eqnarray}

When $\omega_{\lambda} >0$ and we use the rigion of $\omega_{m} <0$,  by take the sum on the integers of $n$ and $m$, the correlation function is calculated as
\begin{eqnarray}
\Phi_{\rm 12}(\omega_{\lambda}) &=& \frac{I^2}{2V^2}\sum_{{\bf k}, {\bf q}} \frac{e\hbar k_x}{m}\frac{2D^2q^2q_x}{\hbar} 2 \int_{-\infty}^{\infty} \frac{n(y)}{2\pi i}  \biggr[ -\bigr[F_{A}({\bf k},{\bf q},y, i\omega_{\lambda}) + F_{B}({\bf k},{\bf q},y, i\omega_{\lambda}) \bigr]D^{A}({\bf q},y)D^{R}({\bf q},y + i\omega_{\lambda}) \nonumber \\
&& + \bigr[F_{A}({\bf k},{\bf q},y-i\omega_{\lambda}, i\omega_{\lambda}) + F_{B}({\bf k},{\bf q},y-i\omega_{\lambda}, i\omega_{\lambda}) \bigr]D^{A}({\bf q},y-i\omega_{\lambda})D^{R}({\bf q},y) \biggr], \nonumber \\
&&
\end{eqnarray}
where $n(y) = 1/(e^{\beta y} -1)$ and $R$ ($A$) indicates the retarded (advanced) Green's function.
$F_{A}({\bf k},{\bf q},y, i\omega_{\lambda})$ and $F_{B}({\bf k},{\bf q},y, i\omega_{\lambda})$ are
\begin{eqnarray}
F_{A}({\bf k},{\bf q},y, i\omega_{\lambda}) &=& -\int_{-\infty}^{\infty}\frac{f(x)}{2\pi i} dx \biggr[ 
 \bigr[ G_{\uparrow}^{R}({\bf k},x) - G_{\uparrow}^{A}({\bf k},x) \bigr] G_{\downarrow}^{R}({\bf k}+{\bf q},x+ i\omega_{\lambda} + y )G_{\uparrow}^{R}({\bf k},x + i\omega_{\lambda}) \nonumber \\
&&  
+G_{\uparrow}^{A}({\bf k},x-y-i\omega_{\lambda})  \bigr[ G_{\downarrow}^{R}({\bf k}+{\bf q},x) - G_{\downarrow}^{A}({\bf k}+{\bf q},x) \bigr] G_{\uparrow}^{R}({\bf k},x-y) \bigr] \nonumber \\   
&&  
+G_{\uparrow}^{A}({\bf k},x - i\omega_{\lambda})  G_{\downarrow}^{A}({\bf k}+{\bf q},x + y) \bigr[ G_{\uparrow}^{R}({\bf k},x) - G_{\uparrow}^{A}({\bf k},x) \bigr] \biggr].\\ 
F_{B}({\bf k},{\bf q},y, i\omega_{\lambda}) &=&  -\int_{-\infty}^{\infty}\frac{f(x)}{2\pi i} dx \biggr[ 
 \bigr[ G_{\downarrow}^{R}({\bf k},x) - G_{\downarrow}^{A}({\bf k},x) \bigr] G_{\uparrow}^{R}({\bf k}-{\bf q},x- y )G_{\downarrow}^{R}({\bf k},x + i\omega_{\lambda}) \nonumber \\
&&  
+G_{\downarrow}^{A}({\bf k},x+y)  \bigr[ G_{\uparrow}^{R}({\bf k}-{\bf q},x) - G_{\uparrow}^{A}({\bf k}-{\bf q},x) \bigr] G_{\downarrow}^{R}({\bf k},x+y+i\omega_{\lambda}) \bigr] \nonumber \\   
&&  
+G_{\downarrow}^{A}({\bf k},x - i\omega_{\lambda})  G_{\uparrow}^{A}({\bf k}-{\bf q},x - y-i\omega_{\lambda}) \bigr[ G_{\downarrow}^{R}({\bf k},x) - G_{\downarrow}^{A}({\bf k},x) \bigr] \biggr].
\end{eqnarray}
where $f(x) = 1/(e^{\beta(x-\mu)} + 1)$ and $G^{R}({\bf k},x)$ is given in Eq. (\ref{GF1}), and $G^{A}= (G^R)^{*}$.
It should be noted that we transformed the variable $x$ as $x+\mu \rightarrow x$.

Using the analytic continuation ($i\omega_n \rightarrow \hbar\omega + i\delta$), and $D^{R}D^{A} \simeq \pi\delta(y - \hbar \omega_{\bf q})/\Gamma_{\rm mag} $, the correlation function becomes
\begin{eqnarray}
\Phi_{\rm 12}(\hbar\omega + i\delta) &=& \frac{I^2}{2V^2}\sum_{{\bf k}, {\bf q}} \frac{e\hbar k_x}{m}\frac{2D^2q^2q_x}{\hbar} 2 \frac{\hbar }{2\pi i} \frac{\pi}{\Gamma_{\rm mag}(T)}\frac{-\beta e^{\beta \hbar \omega_{{\bf q}}}}{(e^{\beta \hbar \omega_{\bf q}}  - 1)^2 } \bigr[ A({\bf k},{\bf q},+, \hbar\omega_{\bf q})+ A({\bf k},-{\bf q},-, -\hbar\omega_{\bf q}) \bigr],
\end{eqnarray}
where $A({\bf k},{\bf q},\sigma, \hbar\omega_{\bf q})$ is given as
\begin{eqnarray}
A({\bf k},{\bf q},\sigma, \hbar\omega_{\bf q}) &=& -\int_{-\infty}^{\infty}\frac{f(x)}{2\pi i} dx \biggr[ 
 \bigr[ G_{\sigma}^{R}({\bf k},x) - G_{\sigma}^{A}({\bf k},x) \bigr] G_{-\sigma}^{R}({\bf k}+{\bf q},x+\hbar\omega_{\bf q})G_{\sigma}^{R}({\bf k},x) \nonumber \\
&&  
+G_{\sigma}^{A}({\bf k},x-\hbar\omega_{\bf q})  \bigr[ G_{-\sigma}^{R}({\bf k}+{\bf q},x) - G_{-\sigma}^{A}({\bf k}+{\bf q},x) \bigr] G_{\sigma}^{R}({\bf k},x-\hbar\omega_{\bf q}) \bigr] \nonumber \\   
&&  
+G_{\sigma}^{A}({\bf k},x)  G_{-\sigma}^{A}({\bf k}+{\bf q},x+\hbar\omega_{\bf q}) \bigr[ G_{\sigma}^{R}({\bf k},x) - G_{\sigma}^{A}({\bf k},x) \bigr] , \biggr].
\end{eqnarray}
where $\sigma = + (-)$ corresponds to $\uparrow$ ($\downarrow$). 

\begin{figure}[h]
\begin{center}
\rotatebox{0}{\includegraphics[angle=-90,width=0.2\linewidth]{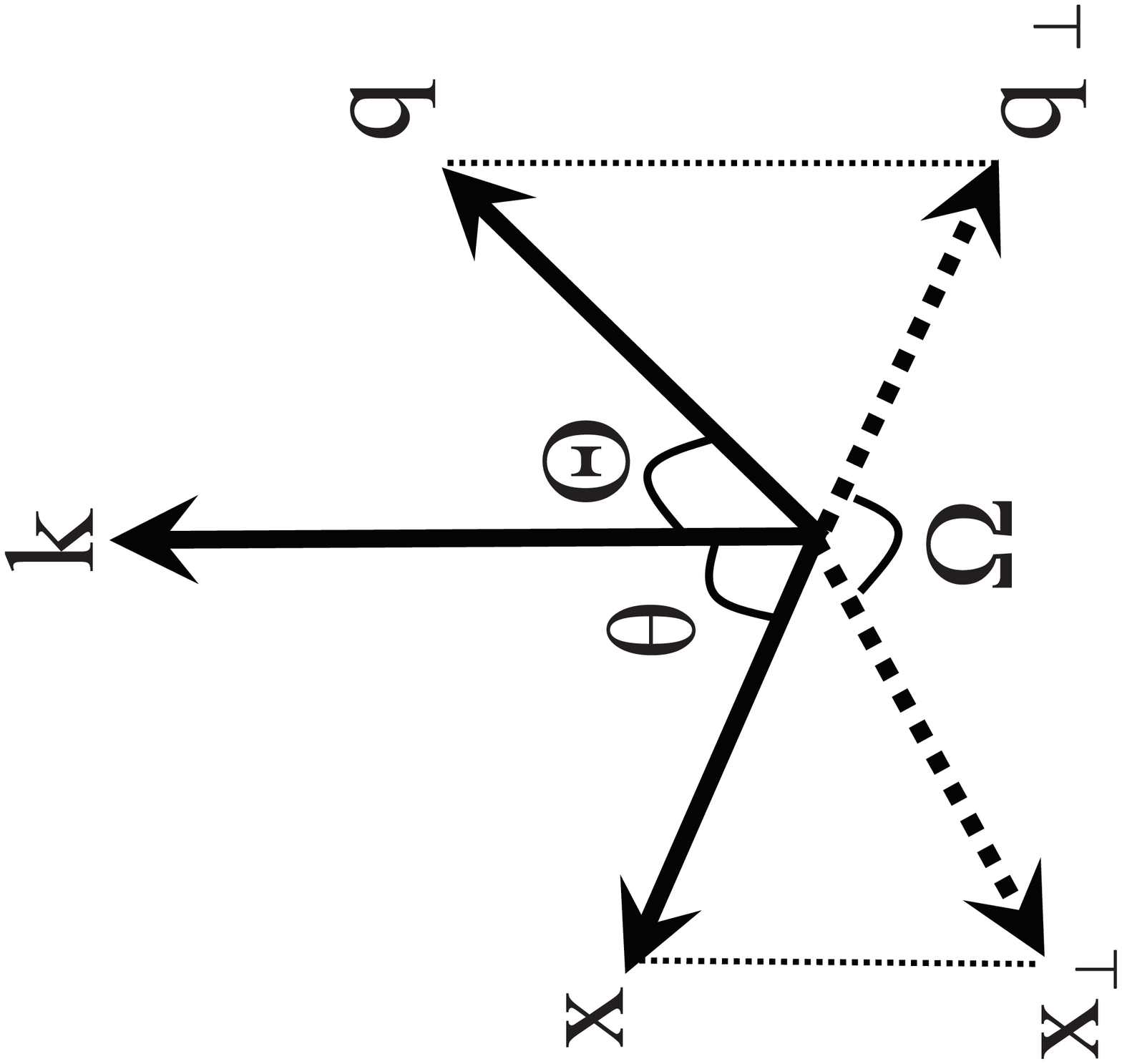}}
\caption{Coordination to perform the integrals. ${\bf q}_{\perp}$ and ${\bf x}_{\perp}$ are the components of ${\bf q}$ and ${\bf x}$ perpendicular to ${\bf k}$.  }
\label{FigA1}
\end{center}
\end{figure}

When we neglect $G^{R}G^{R}G^{R}$ and $G^{A}G^{A}G^{A}$, and we use the following approximations 
\begin{eqnarray}
G^{R}_{\sigma}({\bf k}, x)G^{A}_{\sigma}({\bf k}, x) &=& \frac{\pi}{\Gamma_{\sigma}(x)}\delta(x -\epsilon_{{\bf k}\sigma} - Re\Sigma^{R}_{\sigma}(x)), \\ 
G^{R}_{\sigma}({\bf k}, x) - G^{A}_{\sigma}({\bf k}, x) &=& -2\pi i \delta(x -\epsilon_{{\bf k}\sigma} - Re\Sigma^{R}_{\sigma}(x)),
\end{eqnarray}
$A({\bf k},{\bf q},\sigma, \hbar\omega_{\bf q})$ is approximated as
\begin{eqnarray}
&&A({\bf k},{\bf q},\sigma, \hbar\omega_{\bf q}) \simeq -\int_{-\infty}^{\infty} f(x) dx\biggr[ \frac{\pi}{\Gamma_{\sigma}(x)} \delta(x- \epsilon_{{\bf k}\sigma}  - {\rm Re}\Sigma^{R}_{\sigma}(x) ) \delta(x + \hbar \omega_{\bf q} - \epsilon_{{\bf k} +{\bf q},-\sigma}  -{\rm Re}\Sigma_{-\sigma}^{R}(x + \hbar \omega_{\bf q})) \nonumber \\
&& \hspace{4cm} -  \frac{\pi}{\Gamma_{\sigma}(x-\hbar\omega_{\bf q})} \delta(x- \hbar\omega_{\bf q} - \epsilon_{{\bf k}\sigma}  - {\rm Re}\Sigma^{R}_{\sigma}(x- \hbar\omega_{\bf q}) ) \delta(x - \epsilon_{{\bf k} +{\bf q},-\sigma}  -{\rm Re}\Sigma_{-\sigma}^{R}(x))
\biggr]. 
\end{eqnarray}
To perform the integrals on ${\bf k}$ and ${\bf q}$, we set a new coordination as shown in Fig.\ {\ref{FigA1}}.
In this coordination, $q_x$ and $k_{x}$ are
\begin{eqnarray}
q_x &=& q \cos{\theta} \cos{\Theta} + q\sin{\theta}\sin{\Theta}\cos{\Omega}, \\
k_x &=& k \cos{\theta}.
\end{eqnarray}
By integrating on $k$, $\theta$, $\phi$, $\Theta$ and $\Omega$, the correlation function becomes Eq.\ (7).  

\section{II: Temperature denpendence of Seebeck coefficient and power factor due to the magnon drag: the effect of $\alpha$}
\begin{figure}[h]
\begin{center}
\rotatebox{0}{\includegraphics[angle=-90,width=0.6\linewidth]{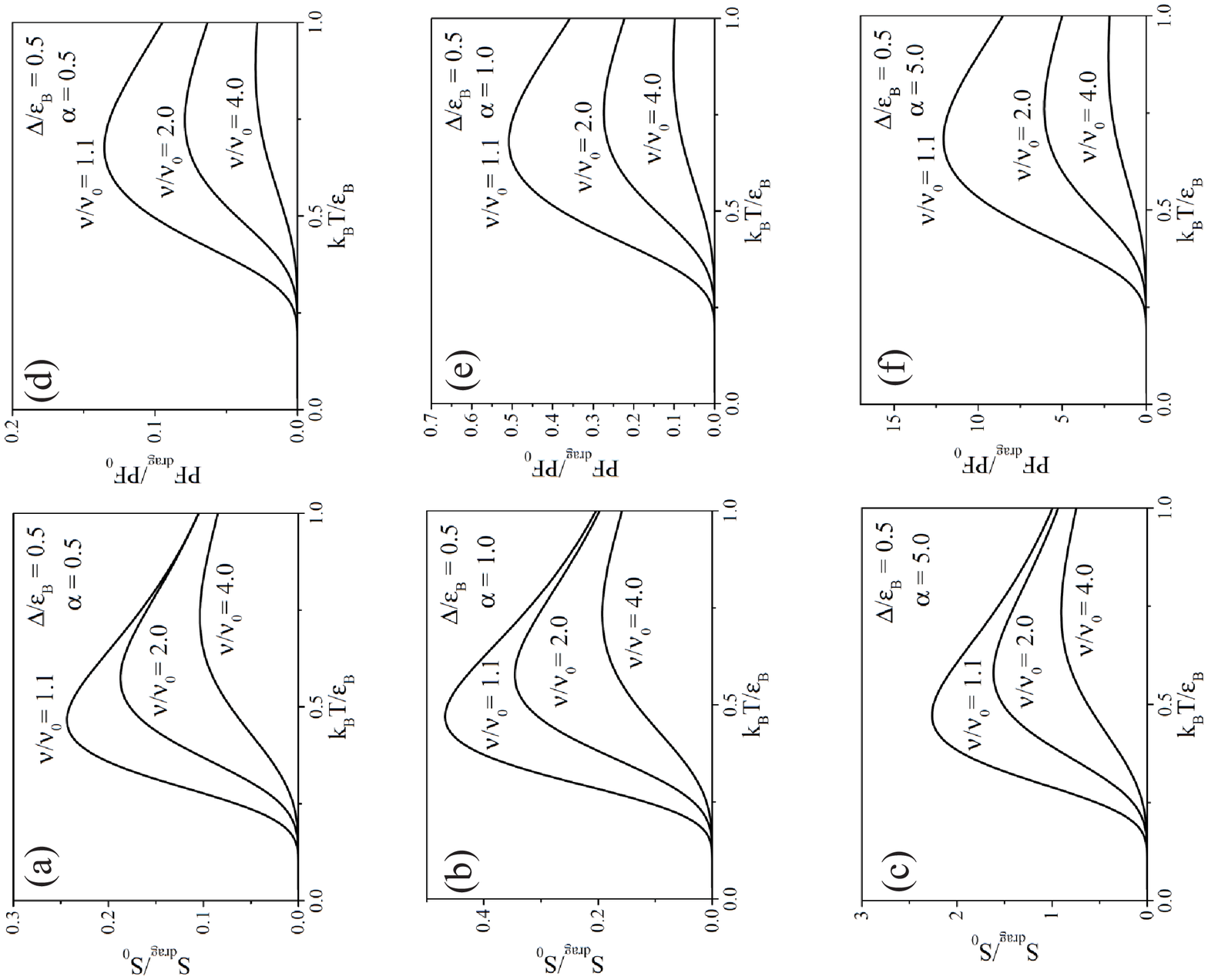}}
\caption{Temperature dependent Seebeck coefficient and power factor due to the magnon drag. We set the parameters as $\alpha = 0.5$ for (a) and (d),  $\alpha =1.0$ for (b) and (d), and $\alpha =5$ for (c) and (e).   }
\label{FigA2}
\end{center}
\end{figure}

Figure \ref{FigA2} shows the temperature dependent Seebeck coefficient and power factor due to the magnon drag. 
The amplitude of $\alpha$ is set as $\alpha = 0.5$ for (a) and (d),  $\alpha =1.0$ for (b) and (d), and $\alpha =5$ for (c) and (e), respectively.
We find that the temperature dependences of Seebeck coefficient and power factor are insensitive to $\alpha$, while the amplitude of Seebeck coefficient and power factor strongly depend on $\alpha$: As $\alpha$ increases, the Seebeck coefficient and the power factor increase.

